\documentclass[aps,pra]{revtex4}
\usepackage{latexsym}
\usepackage{amsmath}
\usepackage{graphicx}
\usepackage{float}

\begin{document}

\title{Exact solution of the extended Hubbard model in the atomic limit on the
Bethe Lattice}
\author{F. Mancini}
\affiliation{ Dipartimento di Fisica "E.R. Caianiello" - Unit\`a
CNISM di Salerno \\Universit\`a degli Studi di Salerno, Via S.
Allende, I-84081 Baronissi (SA), Italy}
\author{F. P. Mancini}
\affiliation{Dipartimento di Fisica and Sezione I.N.F.N.\\
Universit\`a degli Studi di Perugia, Via A. Pascoli, I-06123
Perugia, Italy}
\author{A. Naddeo}
\affiliation{ Dipartimento di Fisica "E.R. Caianiello" - Unit\`a
CNISM di Salerno \\Universit\`a degli Studi di Salerno, Via S.
Allende, I-84081 Baronissi (SA), Italy}

\begin{abstract}
We study the phase diagram at finite temperature of a system of
Fermi particles on the sites of the Bethe lattice with
coordination number $z$ and interacting through onsite $U$ and
nearest-neighbor $V$ interactions. This is a physical realization
of the extended Hubbard model in the atomic limit. By using the
equations of motion method, we exactly solve the model. For an
attractive intersite potential, we find, at half filling, a phase
transition towards a broken particle-hole symmetry state. The
critical temperature, as a function of the relevant parameters,
has a re-entrant behavior as already observed in the equivalent
spin-1 Ising model on the Bethe lattice.
\end{abstract}

\date{\today}
\maketitle

\section{introduction}
\label{sec_1}

Recently, it has been shown \cite{mancini_0506} that a system of
$q$ species of Fermi particles, localized on the sites of a
Bravais lattice, is exactly solvable in any dimension by means of
the equations of motion approach \cite{mancini_04}. This Fermi
system is isomorphic to a spin-$q/2$ Ising model in an external
magnetic field. As a consequence, spin systems can be studied
within a new approach \cite{mancini_0506}. Exactly solvable means
that it is always possible to find a complete set of eigenvalues
and eigenoperators of the Hamiltonian, which closes the hierarchy
of the equations of motion. Thus, one can get exact expressions
for the relevant Green's functions and correlation functions. One
finds that these functions depend on a finite set of parameters to
be self-consistently determined \cite{mancini_04}. It has been
already shown how it is possible to fix such parameters by means
of symmetry and algebra constraints in the case of one dimensional
systems for $q=1,2,3$ \cite{mancini_05}, and in the case of the
Bethe lattice with coordination number $z$ and $q=1$
\cite{mancini_06}.

In this paper, we shall address the case of $q=2$ species of Fermi
particles on the sites of the Bethe lattice with coordination
number $z$ and interacting through nearest-neighbors interaction
$V$. By considering also an onsite interaction $U$, one has the
extended Hubbard model in the atomic limit. We find that for an
attractive nearest-neighbor interaction there is a transition from
a phase where the particle-hole symmetry is preserved to a phase
where this symmetry is broken. The relative variations of the two
regions depend on the coordination number $z$. Furthermore, for
$U/\vert V \vert >2$ the phase diagram exhibits a re-entrant
behavior.

The plan of the paper is as follows: in Sec. \ref{sec_2}, we
outline the equations of motion method for the extended Hubbard
model in the atomic limit. In Sec. \ref{sec_3}, we compute the
Green's and correlation functions and find that they depend only
on two parameters. As a result, in Sec. \ref{sec_4}, we are able
to get a set of self-consistent equations enabling us to fix these
parameters in terms of which all the local properties of the
system can be expressed. In Sec. \ref{sec_5} we shall determine
the transition temperature as a function of the onsite interaction
at half filling and analyze the phase diagram for a wide range of
values of the parameters $T$, $U$ (in units of $\left| V \right|)$
and for different values of the coordination number $z$. In
Appendix A we briefly discuss the equivalence between the extended
Hubbard model in the atomic limit and the spin-1 Ising model on
the Bethe lattice \cite{pawloski_06,katsura_79}. Finally, Sec.
\ref{sec_6} is devoted to our concluding remarks.

\section{The Hamiltonian and the equations of motion}
\label{sec_2}

The extended Hubbard Hamiltonian in the atomic limit is given by
\begin{equation}
\label{eq1} H=-\mu \sum_i n(i)+U\sum_i D(i)+\frac{1}{2}V\sum_{i\ne
j} \,n(i)n(j),
\end{equation}
where $U$ and $V$ represent the onsite and the nearest-neighbor
intersite interaction, respectively. $\mu$ is the chemical
potential, $n(i)=n_\uparrow (i)+n_\downarrow (i)$ and
$D(i)=n_\uparrow (i)n_\downarrow (i)$ are the density and double
occupancy operators at site $i$, respectively. As usual, $n_\sigma
(i)=c_\sigma ^\dag (i)c_\sigma (i)$ with $\sigma =\left\{
{\uparrow ,\downarrow } \right\}$ and $c_\sigma (i)$ ($c_\sigma
^\dag (i))$ is the fermionic annihilation (creation) operator of
an electron of spin $\sigma$ at site $i$, satisfying canonical
anti-commutation relations. In the following we shall use the
spinor notation for all fermionic operators. For instance, $c^\dag
(i)=( {c_\uparrow ^\dag (i)\,\,\,c_\downarrow ^\dag (i)})$. The
sums in Eq. (\ref{eq1}) run over the sites of a Bravais lattice.
Here, we shall consider as Bravais lattice the Bethe lattice,
which is an infinite Cayley tree consisting of a central site -
which we denote by (0) - with $z$ nearest-neighbors forming the
first shell. Each site of a shell is joined to $z-1$
nearest-neighbors to form the second shell, and so on to infinity.
Thus, on the Bethe lattice, the Hamiltonian (\ref{eq1}) may be
conveniently written as
\begin{equation}
\label{eq2}
H=-\mu n(0)+UD(0)+\sum\limits_{p=1}^z H^{(p)},
\end{equation}
where $H^{(p)}$ represents the Hamiltonian of the $p$-th sub-tree
rooted at the central site (0) and it can be written as
\begin{equation}
\label{eq3}
H^{(p)}=-\mu n(p)+UD(p)+Vn(0)n(p)+\sum\limits_{m=1}^{z-1} H^{(p,m)},
\end{equation}
where $p$ are the nearest-neighbors of the site (0). In turns,
$H^{(p,m)}$ describes the $m$-th sub-tree rooted at the site
$(p)$, and so on to infinity.

The density operator cannot be used to employ the standard methods
based on the equations of motion since it does not depend on time.
In order to use the Green's function formalism one may consider
the Hubbard operators $\xi (i)=[n(i)-1]c(i)$ and $\eta
(i)=n(i)c(i)$, obeying to the following equations of motion:
\begin{equation}
\label{eq4}
\begin{split}
 i\frac{\partial }{\partial t}\xi (i)&=-\mu \xi (i)+zV\xi (i)n^\alpha (i) \\
 i\frac{\partial }{\partial t}\eta (i)&=(U-\mu )\eta (i)+zV\eta (i)n^\alpha
(i)\,
 \end{split}
\end{equation}
where $n^\alpha (i)=\sum\nolimits_{p=1}^z n(i,p)/z$, and $(i,p)$
are the nearest neighbors of site $i$. The algebra satisfied by
the operators $n$ and $D$ \cite{mancini_04}, allows one to
establish an important recurrence relation obeyed by $n^\alpha
(i)$:
\begin{equation}
\label{eq5}
[n^\alpha (i)]^k=\sum\limits_{m=1}^{2z} A_m^{(k)} [n^\alpha (i)]^m,
\end{equation}
where the coefficients $A_m^{(k)}$ are rational numbers which can
be easily determined by the algebra and the structure of the Bethe
lattice, and which satisfy the relations $\sum\nolimits_{m=1}^{2z}
A_m^{(k)} =1$ and $A_m^{(k)} =\delta_{m,k}$  $(k=1,\cdots ,2z)$
\cite{mancini_07}. The recurrence relation (\ref{eq5}) is of
seminal importance because it limits the number of composite
operators which can be generated by the dynamics of the original
fermionic operators. In fact, by taking successive time
derivatives of the Hubbard operators $\xi(i)$ and $\eta(i)$, one
clearly sees that for $k>2z$, no additional composite operators
are generated and the equations of motion close
\cite{mancini_0506}. Thus, one may define a new composite field
operator $\psi(i)$ \cite{mancini_04}:
\begin{equation}
\label{eq6}
\psi (i)=\left( {{\begin{array}{*{20}c}
 {\psi ^{(\xi )}(i)} \\
 {\psi ^{(\eta )}(i)} \\
\end{array} }} \right);\quad \psi ^{(\xi )}(i)=\left(
{{\begin{array}{*{20}c}
 {\psi _1^{(\xi )} (i)}  \\
 {\psi _2^{(\xi )} (i)} \\
 \vdots \hfill \\
 {\psi _{2z+1}^{(\xi )} (i)} \\
\end{array} }} \right)=\left( {{\begin{array}{*{20}c}
 {\xi (i)} \hfill \\
 {\xi (i)[n^\alpha (i)]}  \\
 \vdots \\
 {\xi (i)[n^\alpha (i)]^{2z}}  \\
\end{array} }} \right) \quad  \psi ^{(\eta )}(i)=\left(
{{\begin{array}{*{20}c}
 {\psi _1^{(\eta )} (i)}  \\
 {\psi _2^{(\eta )} (i)}  \\
 \vdots  \\
 {\psi _{2z+1}^{(\eta )} (i)} \\
\end{array} }} \right)=\left( {{\begin{array}{*{20}c}
 {\eta (i)}  \\
 {\eta (i)[n^\alpha (i)]}  \\
 \vdots \\
 {\eta (i)[n^\alpha (i)]^{2z}}  \\
\end{array} }} \right).
\end{equation}
By means of the recursion formula (\ref{eq5}), the operators $\psi ^{(\xi )}(i)$
and $\psi ^{(\eta )}(i)$ satisfy the equations of motion:
\begin{equation}
\label{eq7}
\begin{split}
 i\frac{\partial }{\partial t}\psi ^{(\xi )}(i)&=[\psi ^{(\xi
)}(i),H]=\varepsilon ^{(\xi )}\psi ^{(\xi )}(i) \\
 i\frac{\partial }{\partial t}\psi ^{(\eta )}(i)&=[\psi ^{(\eta
)}(i),H]=\varepsilon ^{(\eta )}\psi ^{(\eta )}(i)
 \end{split}
\end{equation}
where $\varepsilon ^{(\xi )}$ and $\varepsilon ^{(\eta )}$ are the
energy matrices of rank $(2z+1)\times (2z+1)$:
\begin{equation}
\label{eq8} \varepsilon ^{(\xi )}=\left( {{\begin{array}{*{20}c}
 {-\mu } & {zV}  & 0 \hfill & \cdots  & 0& 0
 & 0  \\
 0  & {-\mu }  & {zV}  & \cdots  & 0  & 0
 & 0  \\
 0  & 0  & {-\mu }  & \cdots  & 0  & 0
& 0  \\
 \vdots  & \vdots  & \vdots  & \cdots  & \vdots
 & \vdots  & \vdots  \\
 0  & 0  & 0  & \cdots  & {-\mu }  & {zV}
 & 0  \\
 0  & 0  & 0  & \cdots  & 0  & {-\mu }
& {zV}  \\
 0  & {zVA_1^{(2z+1)} }  & {zVA_2^{(2z+1)} }  & \cdots
 & {zVA_{2z-2}^{(2z+1)} }  & {zVA_{2z-1}^{(2z+1)} }
 &
{-\mu +zVA_{2z}^{(2z+1)} }  \\
\end{array} }} \right)
\end{equation}
\begin{equation}
\label{eq9}
\varepsilon ^{(\eta )}=\left( {{\begin{array}{*{20}c}
 {U-\mu }  & {zV}  & 0  & \cdots  & 0  & 0
 & 0  \\
 0  & {U-\mu }  & {zV}  & \cdots  & 0  & 0
 & 0  \\
 0  & 0  & {U-\mu }  & \cdots  & 0  & 0
& 0  \\
 \vdots  & \vdots  & \vdots  & \cdots  & \vdots
 & \vdots  & \vdots  \\
 0  & 0  & 0  & \cdots  & {U-\mu }  & {zV}
 & 0  \\
 0  & 0  & 0  & \cdots  & 0  & {U-\mu }
& {zV}  \\
 0  & {zVA_1^{(2z+1)} }  & {zVA_2^{(2z+1)} }  & \cdots
 & {zVA_{2z-2}^{(2z+1)} }  & {zVA_{2z-1}^{(2z+1)} }  &
{U-\mu +zVA_{2z}^{(2z+1)} }  \\
\end{array} }} \right)
\end{equation}
whose eigenvalues, $E_m^{(\xi )}$ and $E_m^{(\eta )}$, are given
by:
\begin{equation}
\label{eq10}
\begin{split}
E_m^{(\xi )}&=-\mu +(m-1)V \\
E_m^{(\eta)}&=-\mu +U+(m-1)V,
\end{split}
\end{equation}
where $m=1,...,(2z+1)$. It is easy to convince oneself that the Hamiltonian
(\ref{eq2}) has now been formally solved since one has a closed set of
eigenoperators and eigenvalues. Then, by using the formalism of Green's
functions (GF), one can proceed to the calculation of observable quantities.

The two field operators $\psi ^{(\xi )}(i)$ and $\psi ^{(\eta )}(i)$ are
decoupled at the level of equations of motion, as one may clearly see in Eq.
(\ref{eq7}). However, as we shall see in the next Sections, they are coupled via a
set of self-consistent equations allowing for the determination of some
unknown parameters in terms of which observables quantities may be computed.

\section{Retarded Green's function and correlation functions}
\label{sec_3}

The retarded thermal Green's function is defined as:
\begin{equation}
\label{eq11} G^{(s)}(t-{t}')=\langle {R\,\left[ {\psi
^{(s)}(0,t)\,{\psi ^{(s)}}^\dag (0,{t}')} \right]} \rangle =\theta
(t-{t}')\langle {\left\{ {\psi ^{(s)}(0,t),{\psi ^{(s)}}^\dag
(0,{t}')} \right\}} \rangle ,
\end{equation}
where the index $s$ refers either to the Hubbard operator $\xi$ or
$\eta$ and $\langle \cdots \rangle$ denotes the
quantum-statistical average over the grand canonical ensemble. By
means of the field equations (\ref{eq6}), one finds that the
retarded GF satisfies the equation
\begin{equation}
\label{eq12}
\left[ {\omega -\varepsilon ^{(s)}} \right]G^{(s)}(\omega )=I^{(s)},
\end{equation}
where $G^{(s)}(\omega )$ is the Fourier transform of $G^{\left( s
\right)}(t-{t}')$ and $I^{(s)}=\langle {\left\{ {\psi
^{(s)}(0,t),{\psi ^{(s)}}^\dag (0,t)} \right\}} \rangle$ is the
$(2z+1)\times (2z+1)$ normalization matrix. The solution of Eq.
(\ref{eq12}) is \cite{mancini_04}:
\begin{equation}
\label{eq13} G^{(s)}(\omega )=\sum_{m=1}^{2z+1} \frac{\sigma
^{(s,m)}}{\omega -E_m^{(s)} +i\delta },
\end{equation}
where $E_m^{(s)}$ are the eigenvalues of the energy matrices, as
given in Eq. (\ref{eq10}). The spectral density matrices $\sigma
_{ab}^{(s,n)}$ can be computed by means of the formula
\cite{mancini_04}:
\begin{equation}
\label{eq14}
\sigma _{ab}^{(s,n)} =\Omega _{an}^{(s)} \sum\limits_{c=1}^{2z+1} \left[
{\Omega _{nc}^{(s)} } \right]^{-1}I_{cb}^{(s)} .
\end{equation}
In Eq. (\ref{eq14}), $\Omega ^{(s)}$ is the $(2z+1)\times (2z+1)$
matrix whose columns are the eigenvectors of the energy matrix
$\varepsilon ^{(s)}$. It is straightforward to show that $\Omega
^{(\xi )}=\Omega ^{(\eta )}=\Omega$, where the matrix $\Omega$ is
given by:
\begin{equation}
\label{eq15}
\Omega _{p,k} =\left\{
 \begin{array}{cl}
 1  & \quad \quad k=1,\;p=1\\
 0  & \quad \quad k=1,\;p\ne 1\\
(\frac{z}{k-1})^{2z+1-p}  & \quad \quad k\ne 1 \, .
\end{array}
\right.
\end{equation}
%%%%%%%%%
Moreover, the matrix elements of the normalization matrices in Eq.
(\ref{eq14}) can be cast in a simple form as \cite{mancini_05}
\begin{equation}
\label{eq16}
\begin{split}
 I_{n,m}^{(\xi )} &=\kappa ^{(n+m-2)}-\lambda ^{(n+m-2)} \\
 I_{n,m}^{(\eta )} &=\lambda ^{(n+m-2)},\
 \end{split}
\end{equation}
where the charge correlators $\kappa ^{(k)}$ and $\lambda ^{(k)}$ are
defined as
\begin{equation}
\label{eq17}
\begin{split}
 \kappa ^{(k)}&=\langle {[n^\alpha (0)]^k} \rangle \\
 \lambda ^{(k)}&=\frac{1}{2}\langle {n(0)[n^\alpha (0)]^k}
 \rangle .
 \end{split}
\end{equation}
Similarly, one finds that the correlation function (CF)
\begin{equation}
\label{eq18} C^{(s)}(t-{t}')=\langle {\psi ^{(s)}(0,t)\,{\psi
^{(s)}}^\dag (0,{t}')} \rangle =\frac{1}{(2\pi
)}\int\limits_{-\infty }^{+\infty } d\omega \,{\kern
1pt}e^{-i\omega (t-{t}')}C^{(s)}(\omega )\,,
\end{equation}
can also be expressed in terms of the same charge correlators
(\ref{eq17}). In fact, by means of the relation \cite{mancini_06}:
\begin{equation}
\label{eq19}
C(\omega )=-\left[ {1+\tanh \frac{\beta \omega }{2}} \right]\,Im\left[
{G(\omega )} \right],
\end{equation}
where $\beta =1/k_B T$, the CF can be immediately computed from Eq. (\ref{eq13}).
One obtains
\begin{equation}
\label{eq20}
\begin{split}
 C^{(s)}(\omega )&=\pi \,\sum\limits_{m=1}^{2z+1} \sigma ^{(s,n)}T_m^{(s)}
\,\delta \left( {\omega -E_m^{(s)} } \right) \\
 C^{(s)}(t-{t}')&=\frac{1}{2}{\kern 1pt}\,\sum\limits_{m=1}^{2z+1}
e^{-iE_m^{(s)} (t-{t}')}\,\sigma ^{(s,n)}\,T_m^{(s)} ,
 \end{split}
\end{equation}
where $T_m^{(s)} =1+\tanh ( {\beta E_m^{(s)} /2})$. As one can
clearly see, the knowledge of the GF's and, consequently of the
CF's, is not fully achieved. In fact, they depend on the unknown
correlators $\kappa ^{(k)}$ and $\lambda ^{(k)}$which are
expectation values of operators not belonging to the basis
(\ref{eq6}). In the next Section we shall show how these
quantities can be self-consistently computed.

\section{Self-consistent equations}
\label{sec_4}

In order to compute the unknown correlators (\ref{eq17}), one may
start by noticing that, by exploiting the recursion relation
(\ref{eq5}), also $\kappa ^{(k)}$ and $\lambda ^{(k)}$ obey to a
recursion relation which limits their computation just to the
first $2z$ correlators \cite{mancini_07}:
\begin{equation}
\label{eq21}
\begin{split}
 \kappa ^{(k)}&=\sum\limits_{m=1}^{2z} A_m^{(k)} \kappa ^{(m)} \\
 \lambda ^{(k)}&=\sum\limits_{m=1}^{2z} A_m^{(k)} \lambda ^{(m)}.
 \end{split}
\end{equation}
Then, one may split the Hamiltonian (\ref{eq4}) as the sum of two terms:
\begin{equation}
\label{eq22}
\begin{split}
 H&=H_0 +H_I \\
 H_I &=z V n(0)n^\alpha (0).
 \end{split}
\end{equation}
Since $H_0$ and $H_I$ commute, the quantum statistical average of
a generic operator $O$ can be expressed as:
\begin{equation}
\label{eq23} \langle O \rangle =\frac{Tr\{Oe^{-\beta
H}\}}{Tr\{e^{-\beta H}\}}=\frac{\langle {O\,e^{-\beta H_I }}
\rangle _0 }{\langle {e^{-\beta H_I }} \rangle _0 },
\end{equation}
where $\langle \cdots \rangle _0$ stands for the trace with
respect to the reduced Hamiltonian $H_0$. One then considers the
correlation functions
\begin{equation}
\label{eq24} C_{1,k}^{(s)} =\langle {s(0)s^\dag ( 0)\left[
{n^\alpha \left( 0 \right)} \right]^{k-1}}\rangle , \quad s=\xi
,\eta, \quad k=1,...,2z+1 \, ,
\end{equation}
which, by means of Eq. (\ref{eq23}), can be written as:
\begin{equation}
\label{eq25} C_{1,k}^{(s)} =\frac{\langle s(0)s^\dag(0)\left[
n^{\alpha} (0)\right]^{k-1} e^{-\beta H_I }\rangle_0}{\langle
e^{-\beta H_I }\rangle_0}, \quad s=\xi ,\eta , \quad k=1,...,2z+1
\,.
\end{equation}
The Pauli principle leads to the following algebraic relations
\begin{equation}
\label{eq26}
\begin{split}
 \xi^\dag (i)n(i)&=0 \\
 \xi ^\dag (i)D(i)&=0
\end{split} \quad \quad \quad
\begin{split}
\eta ^\dag (i)n(i)&=\eta ^\dag (i) \\
 \eta ^\dag (i)D(i)&=0
\end{split}
\end{equation}
from which one has $\xi ^\dag (0)\,e^{-\beta H_I }=\xi ^\dag (0)$
and $\eta ^\dag (0)\,e^{-\beta H_I }=\eta ^\dag (0)\,e^{-z\beta
Vn^\alpha (0)}$. The Hamiltonian $H_0$ describes a system where
the original lattice has been reduced to the central site (0) and
to $z$ unconnected sublattices. Thus, in the $H_0$-representation,
the correlation functions connecting sites belonging to
disconnected graphs can be decoupled. As a result, Eqs.
(\ref{eq25}) can be rewritten as:
\begin{equation}
\label{eq27} C_{1,k}^{(\xi )} =\frac{\langle {\xi (0)\xi ^\dag
(0)} \rangle _0 \langle {[n^\alpha (0)]^{k-1}} \rangle _0
}{\langle {e^{-\beta H_I }} \rangle _0 },\quad \quad
C_{1,k}^{(\eta )} =\frac{\langle {\eta (0)\eta ^\dag (0)} \rangle
_0 \langle {[n^\alpha (0)]^{k-1}e^{-z\beta Vn^\alpha (0)}} \rangle
_0 }{\langle {e^{-\beta H_I }} \rangle _0 }\,.
\end{equation}
In the $H_0$-representation, the Hubbard operators obey to simple
equations of motion: $[\xi (i),H_0 ]=-\mu \,\xi (i)$ and $[\eta
(i),H_0 ]=-(\mu -U)\,\eta (i)$. Thus, it is easy to show that the
equal time CF's can be expressed as:
\begin{equation}
\label{eq28}
\begin{split}
 \langle {\xi (0)\xi ^\dag (0)} \rangle _0 &=\frac{1}{1+2e^{\beta
\mu }+e^{\beta (2\mu -U)}}=1-B_1 +B_2 \\
 \langle {\eta (0)\eta ^\dag (0)} \rangle _0 &=\frac{e^{\beta \mu
}}{1+2e^{\beta \mu }+e^{\beta (2\mu -U)}}=\frac{1}{2}(B_1 -2B_2 ),
 \end{split}
\end{equation}
where:
\begin{equation}
\label{eq29}
\begin{split}
 B_1 &=\langle {n(0)} \rangle _0 =\frac{2e^{\beta \mu }(1+e^{\beta
(\mu -U)})}{1+2e^{\beta \mu }+e^{\beta (2\mu -U)}} \\
 B_2 &=\langle {D(0)} \rangle _0 =\frac{e^{\beta (2\mu
-U)}}{1+2e^{\beta \mu }+e^{\beta (2\mu -U)}}\,
 \end{split}
\end{equation}
and one has used the identities
\begin{equation}
\label{eq30}
\xi _\sigma \xi _\sigma ^\dag +\eta _\sigma \eta _\sigma ^\dag =1-n_\sigma ,
\quad
\eta _\sigma \eta _\sigma ^\dag =n_\sigma -n_\uparrow n_\downarrow .
\end{equation}
Upon inserting Eqs. (\ref{eq28}) into Eqs. (\ref{eq27}) and by
taking $k=1$, one finds:
\begin{equation}
\label{eq31}
\begin{split}
C_{1,1}^{(\xi )} &=\frac{1-B_1 +B_2}{\langle e^{-\beta H_I}
\rangle_0} \\
C_{1,1}^{(\eta )} &=\frac{\left( {B_1 -2B_2 } \right)\langle
e^{-z\beta V n^\alpha (0)}\rangle_0}{2 \langle e^{-\beta
H_I}\rangle _0}.
\end{split}
\end{equation}
It is not difficult to show that the averages in the above equations can be
expressed as:
\begin{equation}
\label{eq32}
\begin{split}
 \langle {e^{-\beta H_I }} \rangle _0 &=1-B_1 +B_2 +(B_1 -2B_2
)(1+aX_1 +a^2X_2 )^z+B_2 (1+dX_1 +d^2X_2 )^z \\
 \langle {e^{-z\beta Vn^\alpha (0)}} \rangle _0 &=(1+aX_1 +a^2X_2
)^z,
 \end{split}
\end{equation}
where $K=e^{-\beta V}$, $a=(K-1)$, $d=(K^2-1)$. $X_1 $ and $X_2 $
are two parameters defined as:
\begin{equation}
\label{eq33} X_1 =\langle {n^\alpha } \rangle _0, \quad
 \quad X_2 =\langle {D^\alpha } \rangle _0 .
\end{equation}
$X_1$ and $X_2$ are parameters of seminal importance since all
correlators and fundamental properties of the system under study
can be expressed in terms of them. Relevant physical quantities,
such as the mean value of the particle density and doubly
occupancy, and the charge correlators $\kappa^{(k)}$ and
$\lambda^{(k)}$ can be easily computed \cite{mancini_07}:
\begin{equation}
\label{eq34}
\begin{split}
 n&=\frac{(X_1 -2X_2 )(1+aX_1 +a^2X_2 )+2X_2 (1+dX_1 +d^2X_2 )}{(1-X_1 +X_2
)+(X_1 -2X_2 )(1+aX_1 +a^2X_2 )+X_2 (1+dX_1 +d^2X_2 )} \\
 D&=\frac{X_2 (1+dX_1 +d^2X_2 )}{(1-X_1 +X_2 )+(X_1 -2X_2 )(1+aX_1 +a^2X_2
)+X_2 (1+dX_1 +d^2X_2 )},
 \end{split}
\end{equation}
and
\begin{equation}
\label{eq35}
\begin{split}
 \kappa ^{(k)}&=\frac{1}{\langle {e^{-\beta H_I }} \rangle _0
}\left\{ {\langle {[n^\alpha (0)]^k} \rangle _0
+\sum\limits_{m=1}^{2z} (B_1 f_m^{(z)} +B_2 g_m^{(z)} )\langle
{[n^\alpha (0)]^{m+k}} \rangle _0 } \right\} \\
 \lambda ^{(k)}&=\frac{1}{2\langle {e^{-\beta H_I }} \rangle _0
}\left\{ {B_1 \langle {[n^\alpha (0)]^k} \rangle _0
+\sum\limits_{m=1}^{2z} [(B_1 +2B_2 )f_m^{(z)} +2B_2 g_m^{(z)}
]\langle {[n^\alpha (0)]^{m+k}} \rangle _0 } \right\},
 \end{split}
\end{equation}
where $f_m^{(z)}$ and $g_m^{(z)}$ are some easily computable
coefficients depending on the external parameters $T$ and $V$ and
where the expectation value $\langle {[n^\alpha (0)]^k} \rangle _0
$ can also be expressed in terms of $X_1$ and $X_2$
\cite{mancini_07}.

As a result, the solution of the model has been reduced to the
determination of just two parameters. $X_1$ and $X_2$ can be
determined by imposing translational invariance. In particular,
the requests $\langle {n(0)} \rangle =\langle {n^\alpha (0)}
\rangle $ and $\langle {D(0)} \rangle =\langle {D^\alpha (0)}
\rangle $ lead to a set of two self-consistent equations
\cite{mancini_07}:
\begin{equation}
\label{eq36}
\begin{split}
 X_1 &=2e^{\beta \mu }(1-X_1 -dX_2 )(1+aX_1 +a^2X_2 )^{z-1}+e^{\beta (2\mu
-U)}[2+(d-1)X_1 -2dX_2 ](1+dX_1 +d^2X_2 )^{z-1} \\
 X_2 &=e^{\beta (2\mu -U)}[1+dX_1 -(2d+1)X_2 ](1+dX_1 +d^2X_2
)^{z-1}-2e^{\beta \mu }K^2X_2 (1+aX_1 +a^2X_2 )^{z-1}.
 \end{split}
\end{equation}
A full investigation of these equations will be given elsewhere. In this
communication, we shall restrict our analysis to the possibility of a
breakdown of the particle-hole symmetry.

\section{Breakdown of the particle-hole symmetry}
\label{sec_5}

\begin{figure}
\begin{center}
\includegraphics[scale=.75]{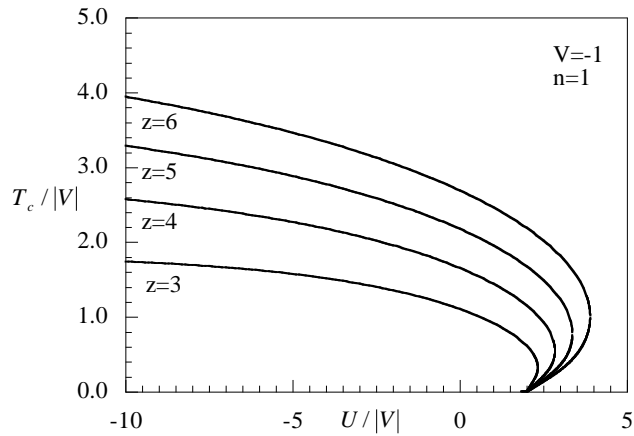}
\caption{\label{fig1} The normalized critical temperature $T_c
/\vert V\vert $ as a function of the on site interaction $U/\vert
V\vert$ for different values of the coordination number $z$ of the
Bethe lattice.}
\end{center}
\vspace{8mm}
\end{figure}

In this Section we shall study the possibility of a spontaneous
breaking of the particle-hole symmetry. Let us consider the
following value of the chemical potential: $\mu =U/2+zV$. For this
value of $\mu$, Eqs. (\ref{eq36}) become
\begin{eqnarray}
\label{eq37} X_1 &=& 2GK^{-z}(1-X_1 -dX_2 )(1+aX_1 +a^2X_2
)^{z-1}+K^{-2z}[2+(d-1)X_1 -2dX_2 ](1+dX_1 +d^2X_2 )^{z-1}\\
 &&  \nonumber\\
 \label{eq38} X_2 &=&K^{-2z}[1+dX_1 -(2d+1)X_2
](1+dX_1 +d^2X_2 )^{z-1}-2GK^{-z+2}X_2 (1+aX_1 +a^2X_2 )^{z-1},
\end{eqnarray}
and Eqs. (\ref{eq34}) become:
\begin{equation}
\label{eq39}
\begin{split}
 n&=\frac{2GK^z(1+aX_1 +a^2X_2 )^z+2(1+dX_1 +d^2X_2 )^z}{K^{2z}+2GK^z(1+aX_1
+a^2X_2 )^z+(1+dX_1 +d^2X_2 )^z} \\
 D&=\frac{(1+dX_1 +d^2X_2 )^z}{K^{2z}+2GK^z(1+aX_1 +a^2X_2 )^z+(1+dX_1
+d^2X_2 )^z},
 \end{split}
\end{equation}
where $G=e^{\beta U/2}$. It is easy to show that for
\begin{equation}
\label{eq40}
X_1 =1-dX_2 \,
\end{equation}
Eq. (\ref{eq37}) is always satisfied. Then, by substituting Eq. (\ref{eq40}) into Eq.
(\ref{eq39}) one obtains
\begin{equation}
\label{eq41} n=1,\quad \quad D=\frac{1}{2+2G(1-a^2X_2 )^z}.
\end{equation}
That is, solution (\ref{eq40}) is in agreement with the particle-hole symmetry.
Upon inserting Eq. (\ref{eq40}) into Eq. (\ref{eq38}) one obtains:
\begin{equation}
\label{eq42}
X_2 (1+K^2)+2GKX_2 (1-a^2X_2 )^{z-1}-1=0.
\end{equation}
Thus, for $\mu =U/2+zV$, Eqs. (\ref{eq36}) admit a solution which
satisfies the particle-hole symmetry and is described by the set
of equations (\ref{eq40}) and (\ref{eq42}). One may ask oneself if
Eqs. (\ref{eq37}) and (\ref{eq38}) do admit other solutions,
different from the one described by (\ref{eq40}) and (\ref{eq42})
and thus breaking the particle-hole symmetry. To this purpose, one
may perturb the solution (\ref{eq40}), by setting $X_1 =1-dX_2
+w$. With a little algebra it is easy to show that a solution with
$w\ne 0$ does exist if, and only if, the following equation
\begin{equation}
\label{eq43}
2K^{\frac{U/\left| V
\right|+2}{2}}[z+K(z-2)]^{z-1}+(K+1)^{z-1}(z-1)^{z-1}[z-K^2(z-2)]=0
\end{equation}
is satisfied. This equation will fix the critical temperature
$T_c$ below which the system admits other solutions, which do not
satisfy the particle-hole symmetry. Numerical calculations show
that Eq. (\ref{eq43}) admits a solution only for negative values
of $V$. Thus, in the following, we shall consider only the case of
attractive intersite potential. It is interesting to consider Eq.
(\ref{eq43}) in two extremal limits, namely $U/\left| V \right|\to
-\infty $ and $T_c \to 0$. In the former limit, as expected, one
finds that the critical temperature is the same as the one of a
spinless fermionic system on the Bethe lattice, equivalent to the
spin-1/2 Ising model \cite{mancini_06,mancini_07,baxter_82}:
\begin{equation}
\label{eq44}
z-K^2(z-2)=0\quad \quad \Rightarrow \quad \quad K_c^2 =\frac{z}{z-2}\quad
\quad \Rightarrow \quad \quad \frac{k_B T_c }{\left| V \right|}=\frac{2}{\ln
\left( {\frac{z}{z-2}} \right)}.
\end{equation}
The limit $T_c \to 0$ provides the value of the ratio $U/\vert
V\vert$ at which one should expect a quantum phase transition:
\begin{equation}
\label{eq45} 2e^{\beta U/2}(z-2)^{z-1}-(z-1)^{z-1}K(z-2)=0\quad
\Rightarrow \quad \left( {U/2-\left| V \right|} \right)=k_B T_c
\ln \left[ {\frac{(z-1)^{z-1}}{2(z-2)^{z-2}}} \right].
\end{equation}
Thus, in the limit $T_c \to 0$, one finds that $U/\vert V\vert
=2$, independently of the value of $z$. The results obtained from
Eq. (\ref{eq43}) are displayed in Fig. \ref{fig1}. One observes
that, at fixed coordination number, by increasing $U$ from large
negative values (i.e., attractive onsite interaction) one finds a
decrease of the critical temperature.

An interesting feature of the model's phase diagram is that it
shows regimes with a re-entrance: namely, fixing $U/\vert V\vert
>2$ and lowering the temperature, one switches from a
particle-hole symmetry preserving phase to another one where this
symmetry is broken at a certain critical temperature. Then, as
evidenced in Fig. \ref{fig1}, lowering further the temperature,
one finds another critical temperature at which the symmetry is
restored. How pronounced the re-entrance is, depends on $z$.
%%%%%%%%%%%
%%%%%%%%%%%%%%%%%%%%%%%%%

\section{Concluding Remarks}
\label{sec_6}

In this paper we have obtained the finite temperature phase
diagram of a system of fermions with onsite and nearest-neighbor
interactions localized on the sites of the Bethe lattice. The
Hamiltonian describing such a system defines the so-called
extended Hubbard model in the atomic limit. Upon using the
equations of motion method, it is possible to exactly solve the
model. For attractive nearest-neighbor interaction we find the
critical temperature at which the system undergoes a transition to
a phase where the particle-hole symmetry is broken. This critical
temperature depends on the ratio $U/\vert V\vert$ and on $z$. It
increases with increasing $z$ and presents a re-entrant behavior
for $U/\vert V\vert >2$.

\appendix
\section{Spin-1 Ising model on the Bethe lattice}

In this Appendix we shall analyze the correspondence between the
extended Hubbard model on the Bethe lattice and the spin-1 Ising
model defined on the same lattice. A transformation from a
fermionic to a spin Hamiltonian can be performed by the use of the
pseudospin variable $S(i)$:
\begin{equation}
S(i)=n_\uparrow (i)+n_\downarrow (i)-1=n(i)-1.
 \label{A.1}
\end{equation}
$S(i)$ can take four values, with $S(i)$=0 double degenerate:
\begin{equation}
\begin{split}
n_\uparrow (i)&=0 \quad \quad \\
n_\uparrow (i)&=1 \quad \quad \\
n_\uparrow (i)&=0 \quad \quad \\
n_\uparrow (i)&=1\quad \quad
\end{split}
\begin{split}
n_\downarrow (i)&=0 \quad \quad \Leftrightarrow\\
n_\downarrow (i)&=0 \quad \quad \Leftrightarrow\\
n_\downarrow (i)&=1 \quad \quad \Leftrightarrow \\
n_\downarrow (i)&=1 \quad \quad \Leftrightarrow\\
\end{split}
\begin{split}
\quad \quad S(i)&=-1 \\
\quad \quad S(i)&=0 \\
\quad \quad S(i)&=0 \\
\quad \quad S(i)&=1
\end{split}
\label{A.2}
\end{equation}
Under the transformation \eqref{A.1} the
Hamiltonian \eqref{eq2} can be cast in the form:
\begin{equation}
\begin{split}
 H&=\Delta \,S^2(0)-h\,S(0)+E_0 +\sum\limits_{p=1}^z H^{(p)} \\
 H^{(p)}&=-J\,S(0)S(p)+\Delta \,\,S^2(p)-h\,S(p)+\sum\limits_{m=1}^{z-1}
H^{(p,m)},
 \end{split}
 \label{A.3}
\end{equation}
where $E_0 =-V+(-\mu +V)N$, $N$ is the total number of sites,
$h=\mu -zV-U/2$, $J=-V $ and $\Delta =U/2$. Thus, the Hamiltonian
\eqref{A.3} appears as the one of a spin-1 Ising model with
nearest-neighbor exchange interaction $J$ in the presence of a
crystal field $\Delta$ and of an external magnetic field $h$. The
difference here is that the Hamiltonian \eqref{A.3} is pertinent
to a four-level system because of the spin degeneracy. It is
possible to get rid of the spin degeneracy by mapping the
fermionic Hamiltonian on the standard spin-1 Ising one with
$\tilde {S}(i)=\left\{ {-1,0,1} \right\}$ paying the price of
making the crystal field $\Delta$ to be temperature dependent
\cite{micnas_84,pawloski_06}: $\tilde {\Delta }=U/2+k_B T\log 2$.
The double degeneracy of every $S(i)=0$ leads to a factor 2 for
every singly occupied site in the partition function of the
classical spin system. This gives rise to an overall factor
\begin{equation}
\prod_i 2^{1-\tilde{S}^2(i)}=2^{\sum_i [1-\tilde{S}^2(i)]}
 \label{A.4}
\end{equation}
One may rewrite the partition function of Hamiltonian \eqref{A.3}
as follows:
\begin{equation}
Z =\sum_{\{S(i)\}=\{-1,0,0,1\}} \exp \{ -\beta H[S(i)]\} =\sum_{\{
\tilde{S}(i)\}=\{-1,0,0,1\}} \exp \{ -\beta \tilde{H}[
\tilde{S}(i)]\}
\end{equation}
where $\tilde{H}$ is the Hamiltonian of the standard spin-1 Ising
model on the Bethe lattice, but now with an effective
temperature-dependent crystal field:
\begin{equation}
\begin{split}
 \tilde {H}&=\tilde {\Delta }\,\tilde {S}^2(0)-h\,\tilde {S}(0)+\tilde {E}_0
+\sum_{p=1}^z \tilde {H}^{(p)} \\
 \tilde {H}^{(p)}&=-J\,\tilde {S}(0)\tilde {S}(p)+\tilde {\Delta }\,\,\tilde
{S}^2(p)-h\,\tilde {S}(p)+\sum_{m=1}^{z-1} \tilde {H}^{(p,m)} \\
 \end{split}
 \label{A.6}
\end{equation}
where $\tilde {E}_0 =E_0 +k_B T\ln 2$ and $\tilde{\Delta}=\Delta
+k_B T\,\ln 2$. Having established the mapping between the two
models, we find that our critical temperature exactly agrees with
the one previously found in the literature \cite{katsura_79}.

\end{document}